\documentclass{aa}  

\usepackage{graphicx}
\usepackage{txfonts}

\usepackage{cleveref}

\usepackage{xcolor}
\begin{document}

   \title{Detection of a warm absorber in a highly variable low-redshift quasar}

   \author{SH. M. Shehata\inst{1,2}
          ,
          Samuzal Barua\inst{3}\fnmsep\thanks{Corresponding author: samuzal.barua@gmail.com}
          \and
          Ranjeev Misra\inst{4}
          }
   \institute{ LUX, CNRS UMR 8262, Observatoire de Paris, 61 Avenue
de l'Observatoire, 75014 Paris, France             
         \and
        National Research Institute of Astronomy and Geophysics, Cairo, Helwan, 11421, Egypt
\and
             Shanghai Astronomical Observatory, Chinese Academy of Sciences, 80 Nandan Road, Shanghai 200030, China          
          \and
          Inter-University Center for Astronomy and Astrophysics, Ganeshkind, Pune, 411007, India
             }

   \date{Received September 15, 1996; accepted March 16, 1997}

  \abstract  
   { 3XMMJ115535.8+232723 (z=0.136) is a low-redshift radio-quiet quasar, which shows an unusual spectral nature. The X-ray properties and variability of this source have never been studied.}
   {Using numerous X-ray observations, this study aims to perform a detailed spectral analysis for 3XMMJ115535.8+232723. Having produced spectra from different flux states, the aims of the study is to investigate long-term spectral behaviour in a single object outside the local group of AGNs.}
   { We used {\it XMM-Newton} and {\it Chandra} observations of quasar 3XMMJ115535.8+232723 obtained between 2000 and 2008 to derive the spectroscopic details. We analysed 0.3--10 keV spectra using an absorbed powerlaw model in combination with a warm absorber component.}
   {The first spectroscopic analysis of 3XMMJ115535.8+232723 reveals that the X-ray spectrum imprints a strong absorption feature, prominent at $\sim 0.7$ keV, which has been interpreted to be due to  absorption in a partially ionized material (a.k.a. warm absorber). The column density and ionisation parameter of the warm absorber vary by more than an order of magnitude for different observations. The results are consistent with high density clouds located ($\sim 1000$ gravitational radius) from the central region and another set of low density clouds further away.  The intrinsic flux and spectral shape of the source also varies significantly over the observations.}
   {}


   \maketitle

%-------------------------------------------------------------------
\section{Introduction}

Warm absorbers have emerged as powerful diagnostics of the central engine of active galactic nuclei (AGN). Identified as partially ionized gas along the line of sight, they imprint characteristic absorption features—most prominently in the soft X-ray regime—shaping the overall spectrum. At least half of Seyfert galaxies, and a smaller fraction of quasars, display warm absorber features \citep[e.g.][]{Reynolds1995, Komossa2000}. Despite their prevalence, their physical origin, location, and variability remain open questions.

The first warm absorber was discovered in the quasar MR~2251-178 (z = 0.064) \citep{Halpern1984}. Subsequent observations with {\it ROSAT} and {\it ASCA} revealed ionized O~VII and O~VIII absorption edges at $\sim$0.74 and $\sim$0.87 keV in many local Seyfert galaxies and quasars \citep[e.g.][]{Nandra1992Natur, Turner1993, Fabian1994, Reynolds1997, Brandt1997b, George1997, George1998, Pan1990, Reeves2000, Krolik2001, Blustin2005, King2012, Psaradaki2025}. {\it XMM-Newton} observations of 21 low-redshift PG quasars also showed that at least half of the objects appear to harbor warm absorbers \citep{Porquet2004}. Further study on quasar MR~2251-178 revealed at least three warm absorbing components with ionisation parameters ranging from ${\rm log}(\xi/{\rm erg\,cm\,s^{-1}}) = 1-3$ \citep{Reeves2013}. In another PG quasar, PG1211+143, the warm absorber was found to cover almost 90\% of the central source, which has a column density $N_{\rm H}= 10^{23.2}$~cm$^{-2}$ and ionisation parameter ${\rm log}(\xi/{\rm erg\,cm\,s^{-1}}) = 1.4$ on average \citep{Papadakis2016}. 

While warm absorbers are commonly detected in nearby AGN, reports of their presence in more distant quasars ($z > 0.1$) are rare. Notable examples include PG~1114+445 ($z = 0.144$) and PG~1309+355 ($z = 0.184$), which show absorption edges with column densities in the range $N_{\rm H}\sim 10^{21}$--$10^{22}$~cm$^{-2}$ \citep{George1997, Ashton2004}. 
Evidence has also been found that the ionized gas shows two warm absorber phases in some quasars, the hot and cold phase determined by the ionisation parameter \citep{Ashton2004}. In the quasar PG\,1114+445, the cooler phase was identified at log ionisation ${\rm log}(\xi/{\rm erg\,cm\,s^{-1}}) = 0.83$, while the hot phase was identified at ${\rm log}(\xi/{\rm erg\,cm\,s^{-1}}) = 2.57$. The same study also presents a single warm absorber phase in the quasar PG\,1309+355 with ${\rm log}(\xi/{\rm erg\,cm\,s^{-1}}) = 1.87$. 

A defining property of warm absorbers is their variability. Changes in column density ($N_{\rm H}$) and/or ionisation parameter (log~$\xi$) provide clues to the structure and dynamics of the circum-nuclear medium. Variability has been observed on timescales ranging from weeks to years.  While variations by factors of ten are relatively common in Seyferts, more extreme changes have also been documented. For example, \citet{Matt2011} reported a factor of $\sim$20 change in column density for NGC 1365 within two months, while other studies have found log~$\xi$ variations of comparable magnitude \citep[e.g.][]{Fabian1994, Silva2018, Korany2019, Sadaula2025, Huang2026}.  Such large-amplitude variations are of particular importance, as they may constrain absorber density, distance, and response timescales. 

However, variability studies have mostly focused on nearby objects, in particular Seyfert galaxies (e.g., NGC\,4051,  RX\,J0134.3-4258 \citep[see]{Komossa2000}). Recently, a moderately distant quasar, 2MASS J14302580+4159572 (z=0.35), has been found to show evidence of variable ionised absorption \citep{Akylas2026}. {\it Swift} monitoring campaign on the quasar PG1211+143 showed that the ionisation parameter varies by a factor of 1.6, which is larger than the average amplitude of the X-ray variations \citep{Papadakis2016}. Warm absorber variability in quasars at higher redshifts ($z > 0.1$) is much less established, largely due to observational limitations: multi-epoch, high signal-to-noise X-ray data are scarce. It remains unclear whether the apparent absence of variability in distant quasars reflects an intrinsic physical difference or simply the lack of adequate observational coverage.

This context makes 3XMMJ115535.8+232723 ($z = 0.136$) a particularly valuable target. Identified in both the 3XMM-DR5 \citep{Rosen2016} and MEXSAS \citep{Vagnetti2016} catalogues, this quasar was previously noted by our group as one of seven sources exhibiting unusually complex spectra not well fit by a simple absorbed power law \citep{shehata2021redshift}. Here we present the first detailed, multi-epoch X-ray spectral study of this source, with the aim of characterizing its warm absorber properties and testing for variability in both $N_{\rm H}$ and ionisation.

\begin{figure}
    \vspace{-0.6cm}
	\includegraphics[width=0.55\textwidth]{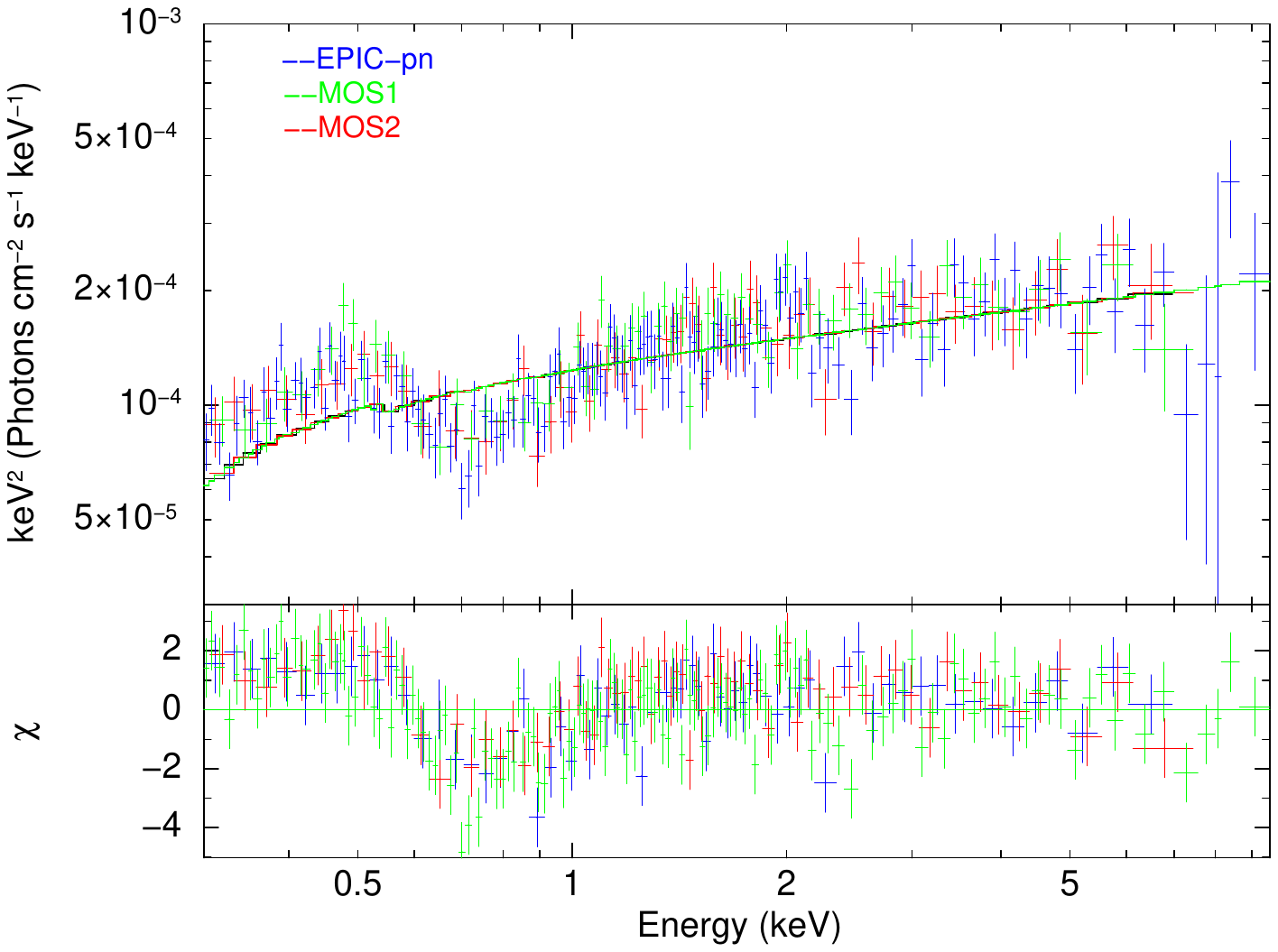} 
	\vspace{-0.5cm}
	\caption{Absorbed powerlaw ({\sc Tbabs$\times$Powerlaw}) fit to the {\em XMM-Newton} spectrum from observation 0502690101. EPIC\,pn, MOS1 and MOS2 spectra are shown in blue, green and red, respectively. Residuals from the fit are presented in the bottom panel, which shows a peak at around 0.7 keV.}
	\label{fig1}
\end{figure}

The paper is constructed as follows: The next section \$2 presents the observations and data analysis. In the following section \$3, we discuss the spectral analysis. The results are discussed and interpreted in section \$4.

\section{Observation and data reduction} \label{sec:observation}

\begin{table*}
\centering
	\caption{{\em XMM-Newton}/EPIC  and {\em Chandra} observations of 3XMMJ115535.8+232723. }
	\begin{tabular}{llllll}
		\hline 
         \hline
		Satellite & Camera & Obs\_ID & Date & Exposure  & Count rate\\
		 & EPIC & & (yyyy-mm-dd) & (ks) & (c/s) \\
		\hline
		%{\em XMM-Newton} & MOS1 & 0112230501 & 2000-12-06 & 25.7 & 1.14E-02 \\ 
		%& MOS2 &  &  & 25.61 & 1.26E-02 \\ 
		{\em XMM-Newton} & pn & 0502690101 & 2007-11-25  & 28.3 & 2.36E-01 \\ 
		& MOS1 &  &  & 41.84 & 6.64E-02 \\ 
		& MOS2 &  &  & 41.8 & 7.24E-02 \\ 
		& pn & 0502690201 & 2007-12-11 & 49.45 & 1.42E-01 \\ 
		& MOS1 &  &  & 63.78 & 4.17E-02 \\ 
		& MOS2 &  &  & 63.82 & 4.53E-02 \\ 
		& pn & 0551280101 & 2008-12-04  & 40.27 & 1.78E-02 \\ 
		& MOS1 &  &  & 55.92 & 5.36E-03 \\ 
		& MOS2 &  &  & 56.06 & 5.39E-03 \\ 
		& pn & 0551280201 & 2008-12-06  & 37.35 & 2.33E-02 \\ 
		& MOS1 &  & & 55.69 & 6.54E-03 \\ 
		& MOS1 &  &  & 55.51 & 7.86E-03  \\ 
		% \hline
		{\em Chandra} & ACIS-I & 537 & 2000-06-23 & 9.561 & 3.91E-02 \\ 
		&  & 1661 & 2001-05-16 & 9.746 & 5.19E-02 \\ 
		&  & 5002 & 2005-02-03 & 36.67 & 2.04E-02 \\ 
		&  & 5003 & 2004-03-06 & 75.05 & 3.91E-02 \\ 
		&  & 7696 & 2007-07-10 & 5.087 & 3.92E-02 \\ 
		\hline
	\end{tabular}
	\label{table1}
\end{table*}

\subsection{{XMM-Newton} data reduction}
Table~\ref{table1} presents the {\it X-ray Multi-Mirror Mission} ({\em XMM-Newton}; \citet{Jansen2001}) observations used in this work. The European Photon Imaging Camera (EPIC) was operated in {\it full window} mode during observations.
We adopted the standard data reduction procedure for {\em XMM-Newton} EPIC PN and MOS by using the Science Analysis System (sas v.20.0.0). {\sc evselect} task was performed to create the event file lists, which were screened and cleaned for high background and any flaring events. Following this, Good Time Interval (GTI) files were created again using {\sc evselect} to make an optimal cut to the background flare. The out-products for the cleaned event files were used to extract the source and background regions. We selected an optimum extraction region for the source, which is 20 arcseconds in radius, whereas another circular region of 40 arcseconds in radius was extracted from the source-free region. The choice of these regions was to maximise the signal-to-noise ratio (S/N) of the data. In the next stage, these regions were generally used to obtain the required data products in the analysis, which include the source spectra, background spectra, source lightcurve and background lightcurve. We further generated rmf (response matrix files) and arf (auxiliary response files) by performing {\sc SAS} tasks {\sc rmfgen} and {\sc arfgen}, respectively. These response files are typically used to binning the data in order to oversample instrumental resolution in energy. Finally, the spectra were grouped using the {\sc specgroup} command.

\begin{figure*}
    %\centering
    \vspace{-1.5cm}
    \includegraphics[width=0.45\textwidth, angle=-90]{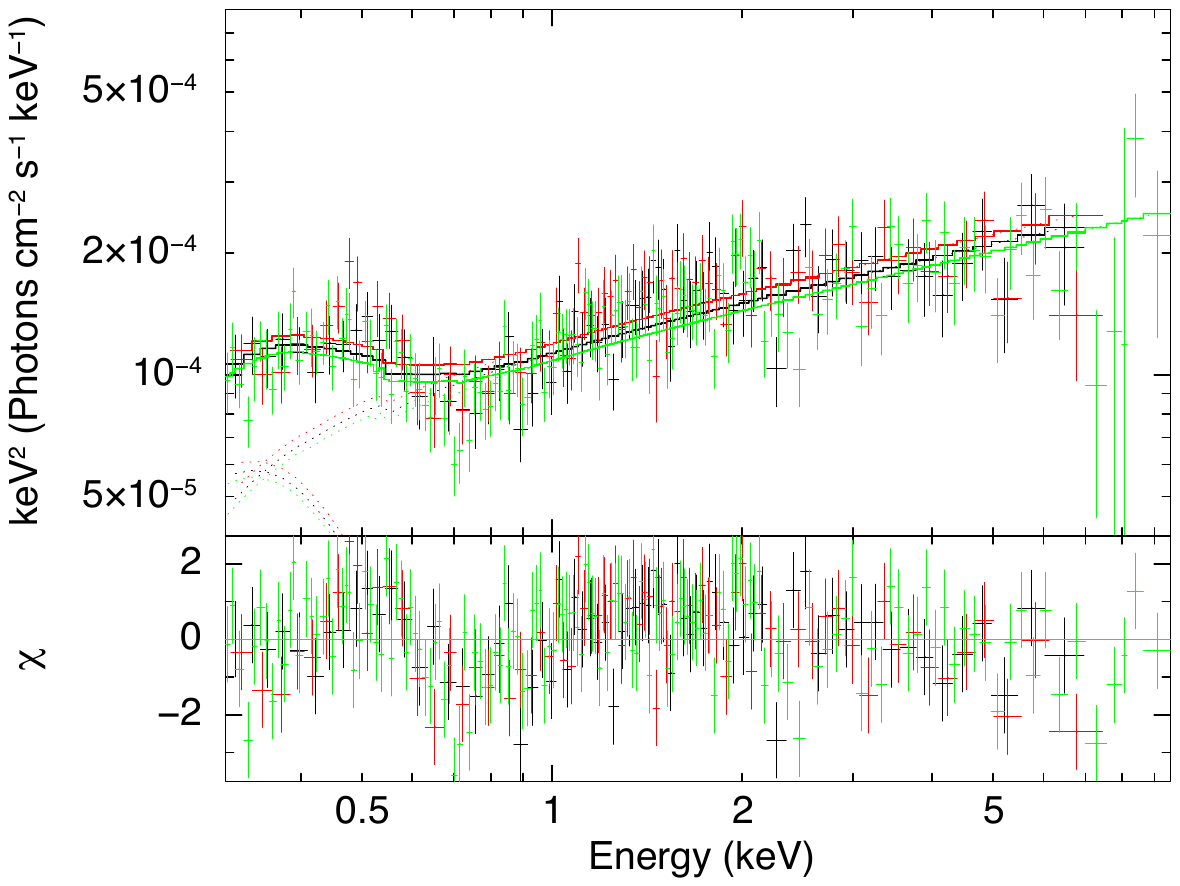}
    \hspace{-2cm}
	\includegraphics[width=0.45\textwidth, angle=-90]{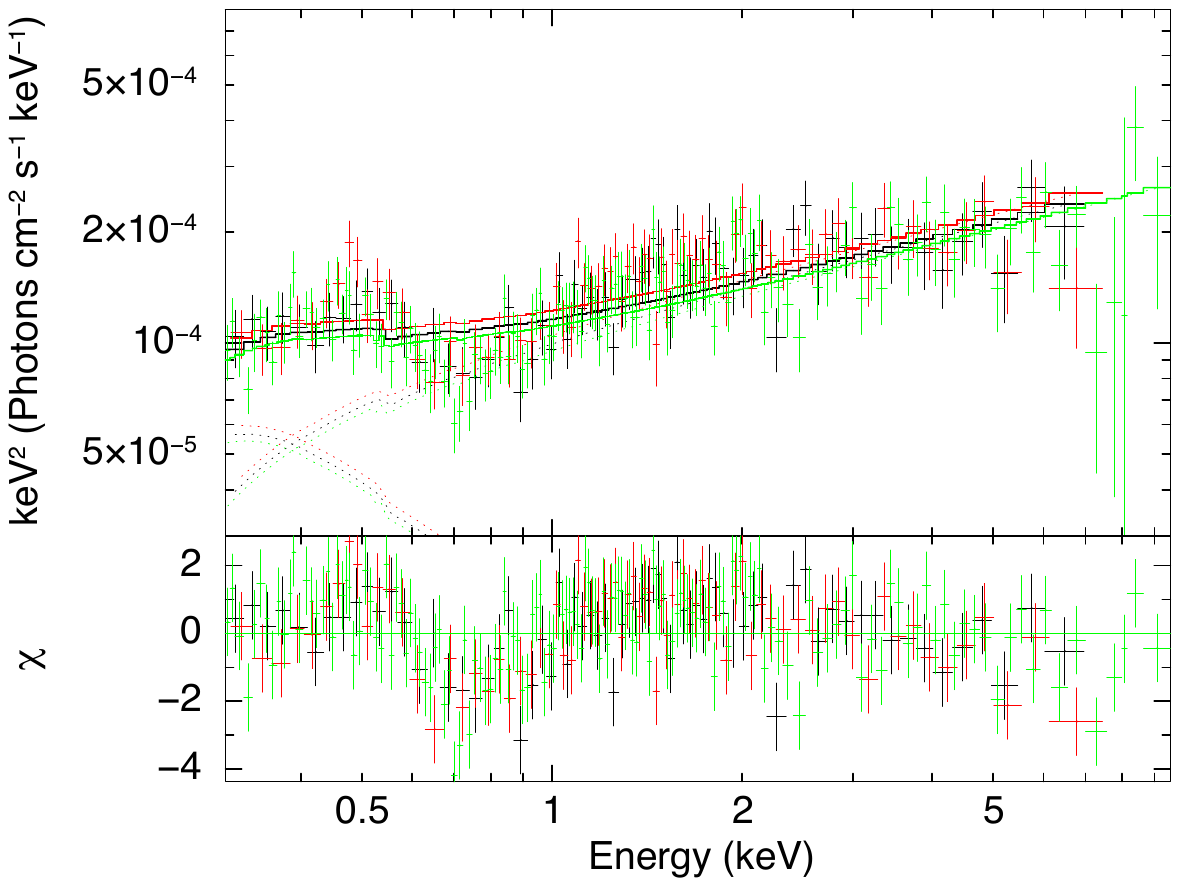} 
	\vspace{-0.5cm}
	\caption{{\em XMM-Newton} spectra from observation 0502690101 fitted using the model {\sc Tbabs $\times$(bbody+Powerlaw)} (left panel) and {\sc Tbabs $\times$(compTT+Powerlaw)} (right panel). EPIC\,pn, MOS1 and MOS2 spectra are shown in blue, green and black, respectively. Spectral fit residuals appear to be persistent in the soft X-ray band below 1 keV.}
	\label{bb+comtt+po_fit}
\end{figure*}

\subsection{{Chandra} data reduction}

{\em Chandra} X-ray observatory \citep{Fruscione2006} performed observations of 3XMMJ115535.8+232723 in other five epochs between 2000 and 2007.
The standard data reduction technique was applied to {\em Chandra} observations as instructed by {\it Chandra X-ray Centre (CXC)}\footnote{https://cxc.cfa.harvard.edu/ciao/guides/}.
Table~\ref{table1} lists the {\it Chandra} observations used in this work. The software required to reduce the data is {\sc chandra interactive analysis of observations (CIAO)} (software version 4.16; ciao-4.16) \citep{Fruscione2006}. With the calibration database (CALDB)\footnote{https://cxc.cfa.harvard.edu/caldb/downloads/index.html} (version 4.9.8.) we reprocessed the level-1 event files using the {\tt chandra\_repro} script. This in turn produced the level-2 event files, which employ the latest calibration and are cleaned for Good Time Intervals (GTI) and eradicating any bad pixel events. Using these clean event files, we extracted the source and background regions by selecting circular apertures of radii of 18 arcseconds. These regions are generally required for doing further analysis and to have the final products. Following this, the {\tt specextract} command was employed to produce the source and background spectra from the respective regions. The spectra were grouped using the {\tt grppha} tool to be used in the spectral analysis.\\

\begin{table*}[!ht]
\centering
\caption{Results from the spectral fit to the {\it XMM-Newton} dataset.}
\begin{tabular}{llllll}
\hline
{\sc Tbabs $\times$ fabsori $\times$ (bbody+Powerlaw)} & \multicolumn{4}{c}{{\em XMM-Newton}} \\
& {0502690101} & {0502690201} & {0551280101} & {0551280201} \\
\hline

$N_{\rm H}\times10^{22}$ (cm$^{-2}$) & $0.35^{+0.19}_{-0.07}$ & $0.36^{+0.29}_{-0.05}$ & $11.73^{+6.58}_{-2.73}$ & $17.39^{+0.80}_{-1.76}$ \\\\

$\xi$ (erg cm s$^{-1}$) & $2.45^{+2.22}_{-1.17}$ & $2.14^{+3.19}_{-2.14}$ & $117.91^{+53.6}_{-66.4}$ & $84.29^{+3.69}_{-3.24}$ \\\\

$C_{f}$ & $\ge0.79$ & $\ge0.72$ & $0.76^{+0.04}_{-0.00}$ & $0.82^{+0.02}_{-0.02}$ \\\\

$kT_{\rm BB}$ (keV) & $0.09^{+0.02}_{-0.01}$ & $0.08^{+0.02}_{-0.02}$ & $0.11^{+0.20}_{-0.10}$ & $0.05^{+0.06}_{-0.04}$ \\\\

$A_{\rm BB}$ ($\times10^{-6}$) & $1.96^{+0.94}_{-0.86}$ & $1.53^{+13.65}_{-0.86}$ & $0.04^{+0.03}_{-0.03}$ & $0.38^{+0.23}_{-0.00}$ \\\\

$\Gamma$ & $1.98^{+0.07}_{-0.07}$ & $1.92^{+0.09}_{-0.08}$ & $1.50^{+0.20}_{-0.00}$ & $1.50^{+0.01}_{-0.00}$\\\\

$A_{\rm PL}$ ($\times10^{-4}$) & $1.78^{+0.17}_{-0.14}$ & $1.10^{+0.13}_{-0.09}$ & $0.28^{+0.09}_{-0.03}$ & $0.38^{+0.01}_{-0.00}$\\\\

$F_{\rm 2-10~keV}~\times 10^{-12}$ {\rm( erg $\rm cm^{-2} s^{-1}$)} & $0.43 \pm 0.01$ & $0.30 \pm 0.01$ & $0.12 \pm 0.05$  & $0.16 \pm 0.05$ \\\\

$F_{\rm 0.3-10~keV}~\times 10^{-12}$ {\rm( erg $\rm cm^{-2} s^{-1}$)} & $0.78 \pm 0.01$ & $0.51 \pm 0.01$ & $0.14 \pm 0.03$  & $0.18 \pm 0.03$ \\\\

$L_{\rm 2-10~keV}~\times 10^{43}$ {\rm( erg $\rm  s^{-1}$)} & $2.13 \pm 0.01$ & $1.55 \pm 0.02$ & $0.82 \pm 0.07$  & $1.23 \pm 0.07$ \\\\

$L_{\rm 0.05-300~keV}~\times 10^{43}$ {\rm( erg $\rm  s^{-1}$)} & $8.61 \pm 0.02$ & $4.99 \pm 0.02$ & $2.14 \pm 0.15$  & $2.36 \pm 0.07$ \\\\
\hline

$\chi^{2}$/{\rm d.o.f.} & 270/290 & 218/206 & 48/39 & 51/47 \\

\hline
\end{tabular}
\label{tab:xmm_fabsori_bb}

\tablefoot{The spectra are modelled with the {\sc Tbabs} $\times$ {\sc fabsori} $\times$ ({\sc bbody} + {\sc powerlaw}) model. From top to bottom, the parameters are the column density ($N_{\rm H}$) and ionisation parameter ($\xi$) of the partially covering absorber, its covering fraction ($C_{f}$), the blackbody temperature ($kT_{\rm BB}$) and normalization ($A_{\rm BB}$), the photon index ($\Gamma$) and normalization ($A_{\rm PL}$) of the power law, the observed 2--10 keV and 0.3--10 keV fluxes, the intrinsic 2--10 keV and 0.05--300 keV luminosities, and the reduced $\chi^{2}$ per degrees of freedom (d.o.f.). The quoted error bars are at the 90\% confidence level.}

\end{table*}

\begin{figure*}
\vspace{-0.8cm}
\hspace{-1.0cm}
\includegraphics[width=0.5\textwidth, angle=-90]{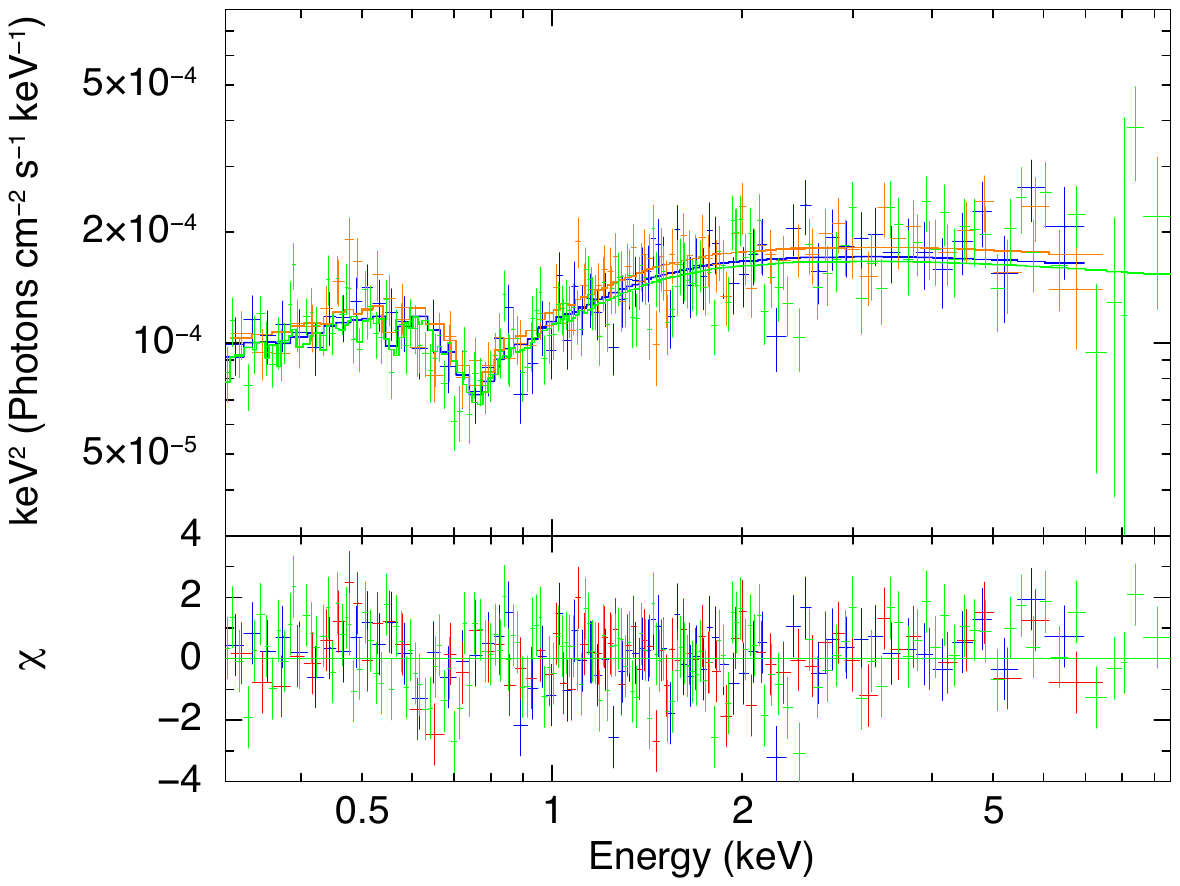}
\hspace{-2.5cm} 
\vspace{-2cm}
\includegraphics[width=0.5\textwidth, angle=-90]{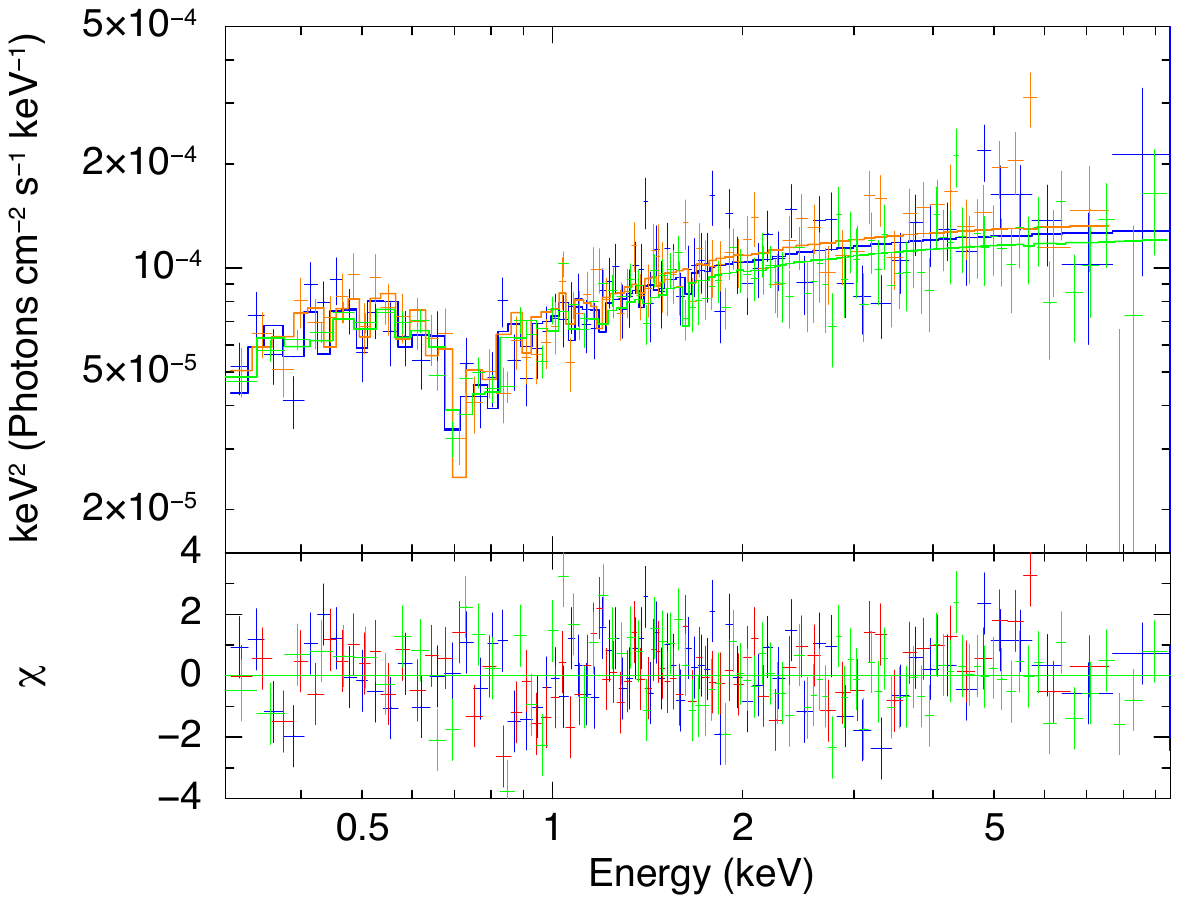}
\vspace{0cm}
\hspace{-1cm}
\includegraphics[width=0.5\textwidth, angle=-90]{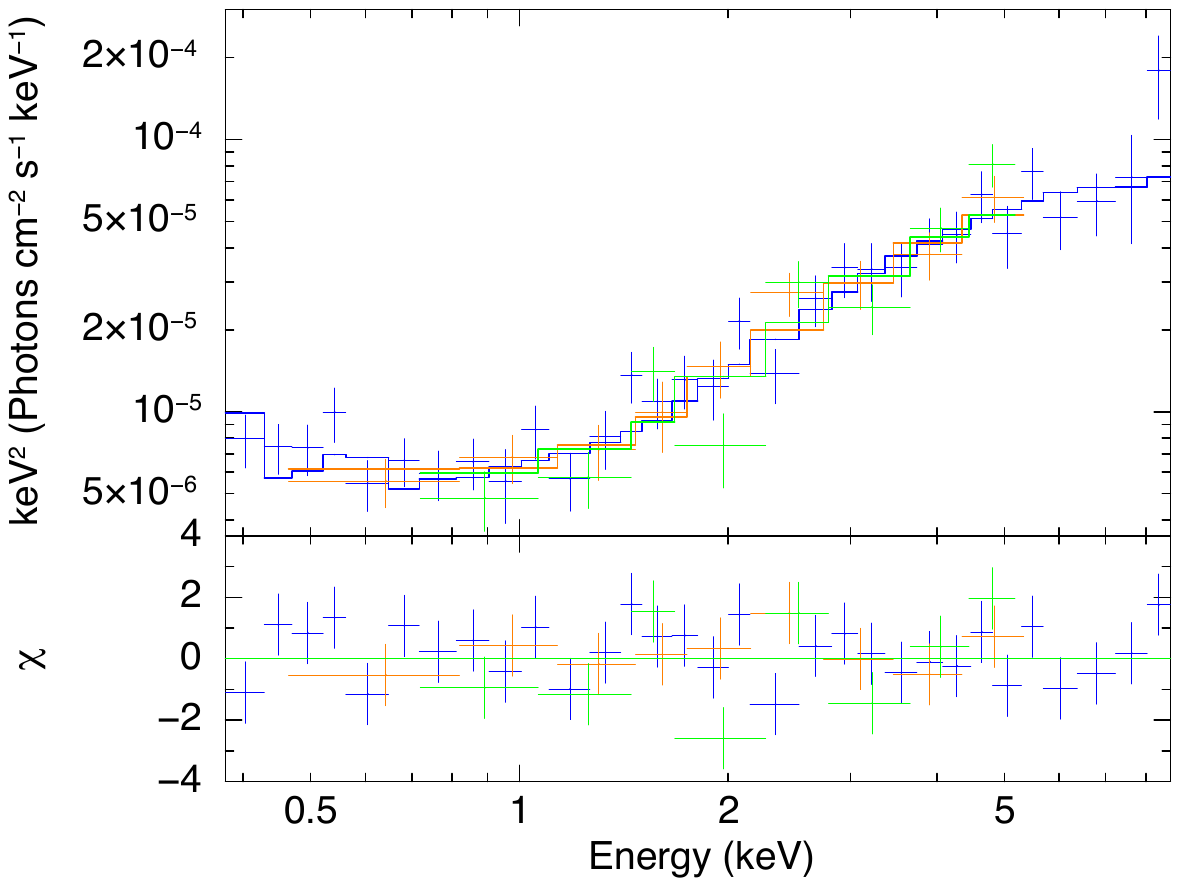} 
\hspace{-2.5cm}
\includegraphics[width=0.5\textwidth, angle=-90]{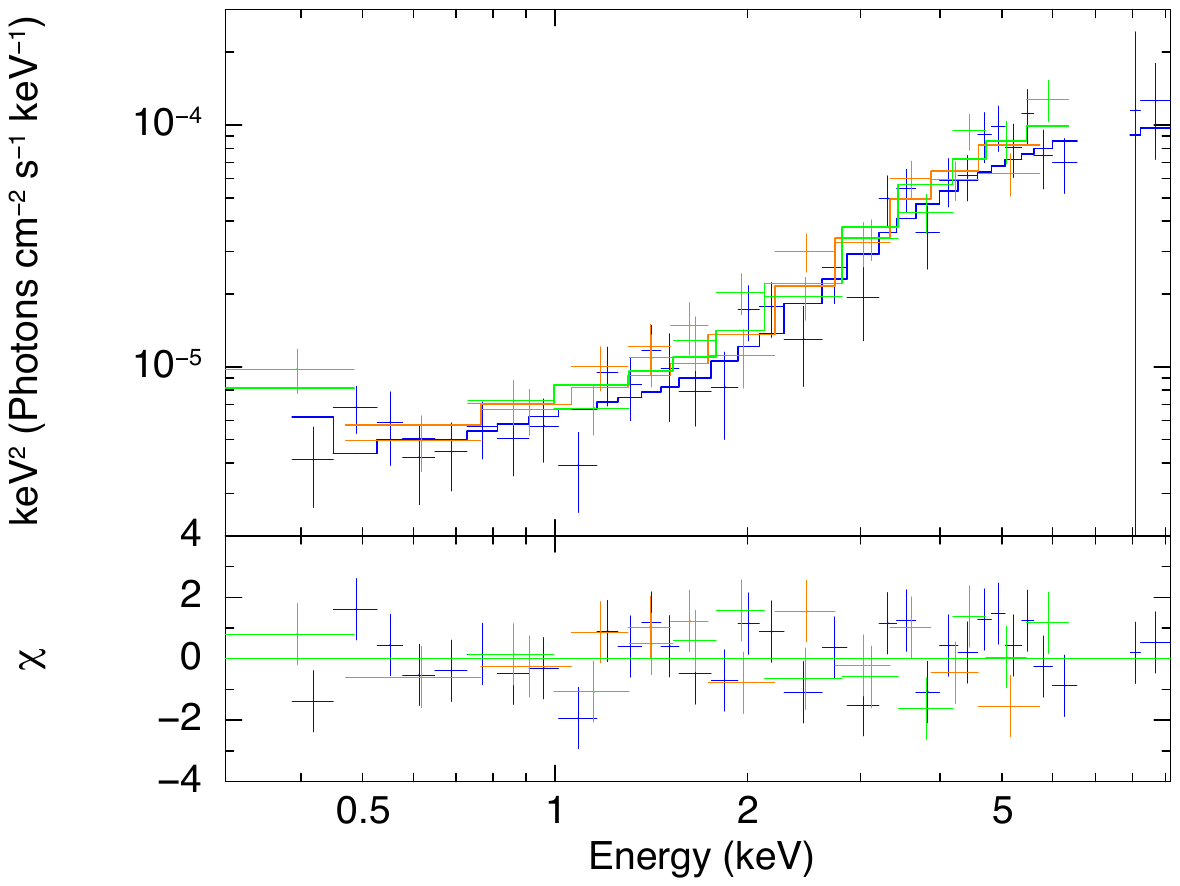}  
\caption{Representative spectra obtained from {\em XMM-Newton} EPIC PN ({\it  Orange}), MOS1 ({\it green}) and MOS2 ({\it blue}) cameras produced from observations 0502690101 (top left), 0502690201 (top right), 0551280101 (bottom left) and 0551280201 (bottom right). Unfolded spectra are shown to indicate the presence absorption feature at 0.7 keV. The spectra fits are jointly performed using the model combination {\sc Tbabs $\times$ absori $\times$ Powerlaw}.}
\label{fig3}
\end{figure*}

\begin{table*}[!ht]
\centering
\caption{Best-fit parameters from the joint spectral fitting of the {\it Chandra} observations.}
\begin{tabular}{lllllll}
\hline
{\sc Tbabs $\times$ fabsori $\times$ (bbody+Powerlaw)} & \multicolumn{4}{c}{{\em Chandra}}\\
& {1661} & {5002} & {5003} & {537+7696} \\
\hline
$N_{\mathrm{H}} \times 10^{22}$ (cm$^{-2}$) & $0.79^{+2.39}_{-0.12}$ & $2.74^{+0.04}_{-0.03}$ & $0.23^{+0.42}_{-0.07}$ & $0.38^{+0.66}_{-0.00}$ \\\\

$\xi$ (erg cm s$^{-1}$) & $0.09^{+0.69}_{-0.08}$ & $0.12^{+0.01}_{-0.01}$ & $4.75^{+20.04}_{-3.12}$ & $0.54^{+3.32}_{-0.00}$ \\\\

$C_f$ & $\ge0.63$ & $0.68^{+0.02}_{-0.01}$ & $\ge0.55$ & $\ge0.73$ \\\\

$kT_{\rm BB}$ (keV) & $0.14^{+0.03}_{-0.10}$ & $0.63^{+0.02}_{-0.03}$ & $0.06^{+0.02}_{-0.02}$ & $0.06^{+0.00}_{-0.01}$ \\\\

$A_{\rm BB}$ ($\times10^{-6}$) & $15.98^{+10.84}_{-10.07}$ & $0.15^{+0.50}_{-0.02}$ & $5.78^{+9.78}_{-4.22}$ & $2.52^{+5.13}_{-0.76}$\\\\

$\Gamma$ & $1.50^{+1.27}_{-0.00}$ & $1.72^{+0.13}_{-0.10}$ & $1.68^{+0.14}_{-0.10}$ & $1.72^{+0.36}_{-0.00}$ \\\\

$A \times 10^{-4}$ & $1.14^{+1.41}_{-0.00}$ & $0.81^{+0.01}_{-0.01}$ & $0.83^{+0.18}_{-0.10}$  & $1.25^{+0.65}_{-0.00}$ \\\\

$F_{\rm 2-10~keV}~\times 10^{-12}$ {\rm( erg $\rm cm^{-2} s^{-1}$)} & $ 0.56 \pm 0.03$ & $0.29 \pm 0.03$ & $0.33 \pm 0.02$  & $0.46 \pm 0.04 $ \\\\

$F_{\rm 0.3-10~keV}~\times 10^{-12}$ {\rm( erg $\rm cm^{-2} s^{-1}$)} & $ 0.78 \pm 0.03$ & $0.59 \pm 0.03$ & $0.53 \pm 0.02$  & $0.79 \pm 0.04 $  \\\\

$L_{\rm 2-10~keV}~\times 10^{43}$ {\rm( erg $\rm  s^{-1}$)} & $2.70 \pm 0.07$ & $1.61 \pm 0.05$ & $1.66 \pm 0.03$  & $2.19 \pm 0.09$ \\\\

$L_{\rm 0.05-300~keV}~\times 10^{43}$ {\rm( erg $\rm  s^{-1}$)} & $4.33 \pm 0.07$ & $2.66 \pm 0.05$ & $3.18 \pm 0.03$  & $4.07 \pm 0.05$  \\\\

$\chi^2/{\mathrm{d.o.f.}}$ & 32/23 & 39/38 & 156/141 & 31/27 \\
\hline
\end{tabular}
\label{tab:chandra_fabsorbi_bb}
\tablefoot{The columns correspond to the individual {\it Chandra} observation IDs, with the final column representing the joint fit to the 537 and 7696 datasets. Parameters from top to bottom are sequentially: the column density ($N_{\rm H}$) and ionisation parameter ($\xi$) of the partially covering absorber; the covering fraction ($C_f$); the blackbody temperature ($kT_{\rm BB}$) and normalization ($A_{\rm BB}$); the photon index ($\Gamma$); the power-law normalization ($A_{\rm PL}$); the observed 2--10 keV and 0.3--10 keV fluxes ($F$); the intrinsic 2--10 keV and 0.05--300 keV luminosities ($L$); and the reduced $\chi^2$ per degrees of freedom. The quoted errors are at the 90\% confidence level.}
\end{table*}

\section{X-ray spectral Analysis}

We started the spectral fitting procedure by using a simple powerlaw model plus galactic absorption to the multi-epoch {\it XMM-Newton} spectra. Spectra from both EPIC pn/MOS instruments were jointly analysed. 
For galactic absorption, we used the {\sc xspec} model {\sc Tbabs} \citep{Wilms2000}, with a column density\footnote{https://www.swift.ac.uk/analysis/nhtot/index.php} $N_{\rm H} = 1.7 \times 10^{20}$ $\rm cm^{-2}$. 
The simple model, however, could not provide a satisfactory fit to the spectra, providing a reduced $\chi^2 /d.o.f. = 507/295$, obtained from the {\it XMM-Newton} observation 0502690101. This observation is just shown as a representative of all other observations used. The resulting spectrum from {\it XMM-Newton} EPIC pn/MOS is produced in Fig.\,\ref{fig1}, which shows a significant absorption feature at $\sim0.7$ keV.
Spectral analysis was performed in {\sc xspec} (version: 12.11.1) \citep{Arnaud1996}. We estimated the absorbed luminosity $L_{\rm X}$ using the {\sc xspec} model {\sc clumin} and flux $F_{\rm X}$ using the {\sc cflux} model to observe the long-term X-ray variability between observations.
The errors were estimated at a confidence level of 90\%.

Next, we investigated whether the blackbody component could account for the observed spectra by adding the blackbody model {\sc bbody} to the {\sc powerlaw}. The inclusion of the blackbody model improved the fit relative to the absorbed power-law model, with the residuals becoming reduced; however, further fitting was still required to achieve a satisfactory description of all spectra, as residuals remained below 1 keV. With the {\tt bbody} component included, the spectral fit yielded a reduced $\chi^2$/d.o.f. = 434/293 for the {\it XMM-Newton} observation 0502690101. In Figure~\ref{bb+comtt+po_fit}, we show the fitted spectrum of this observation as representative of all spectra obtained with the model {\sc Tbabs $\times$ (bbody + Powerlaw)}.

Subsequently, we included the ionizer absorber model {\sc absori} \citep{1Done1992, Magdziarz1995} in the existing model {\sc Tbabs $\times$ (bbody + Powerlaw)}. The {\sc absori} model describes the absorption X-ray spectra by a photo-ionized gas in thermal equilibrium. It is often used in the study of AGN and X-ray binaries to model the warm absorber spectra in {\sc xspec}. We used the {\sc fabsori} version of the model, which includes the partial covering fraction as a free parameter, assuming that the absorber only covers some fraction of the source, while the remaining of the spectrum is seen directly. The model combination applied to the entire 0.3--10 keV spectra is now   {\sc Tbabs $\times$ fabsori $\times$ (bbody+Powerlaw)}. In the fit, the blackbody temperature $kT_{\rm BB}$ of the {\sc bbody} component is restricted to the physically motivated range  $<1~ \rm keV$. We tied the photon index between the {\sc fabsori} and {\sc Powerlaw} components, while freezing the iron abundance to the default value of 1. The other {\sc fabsori} parameters, including column density $N_{\rm H}$, ionisation $\xi$, and the covering fraction $C_{f}$, were allowed to vary independently. Inclusion of the warm absorber significantly improved the fit,  flattening the residuals below 1 keV.
The fit parameters from the {\it XMM-Newton} spectra are presented in Table~\ref{tab:xmm_fabsori_bb}.  
The best-fit {\em XMM-Newton} spectra are shown in Fig.\,\ref{fig3}.

\begin{figure*}
\vspace{-1.85cm}
\hspace{-1.0cm}
\includegraphics[width=0.45\textwidth, angle=-90]{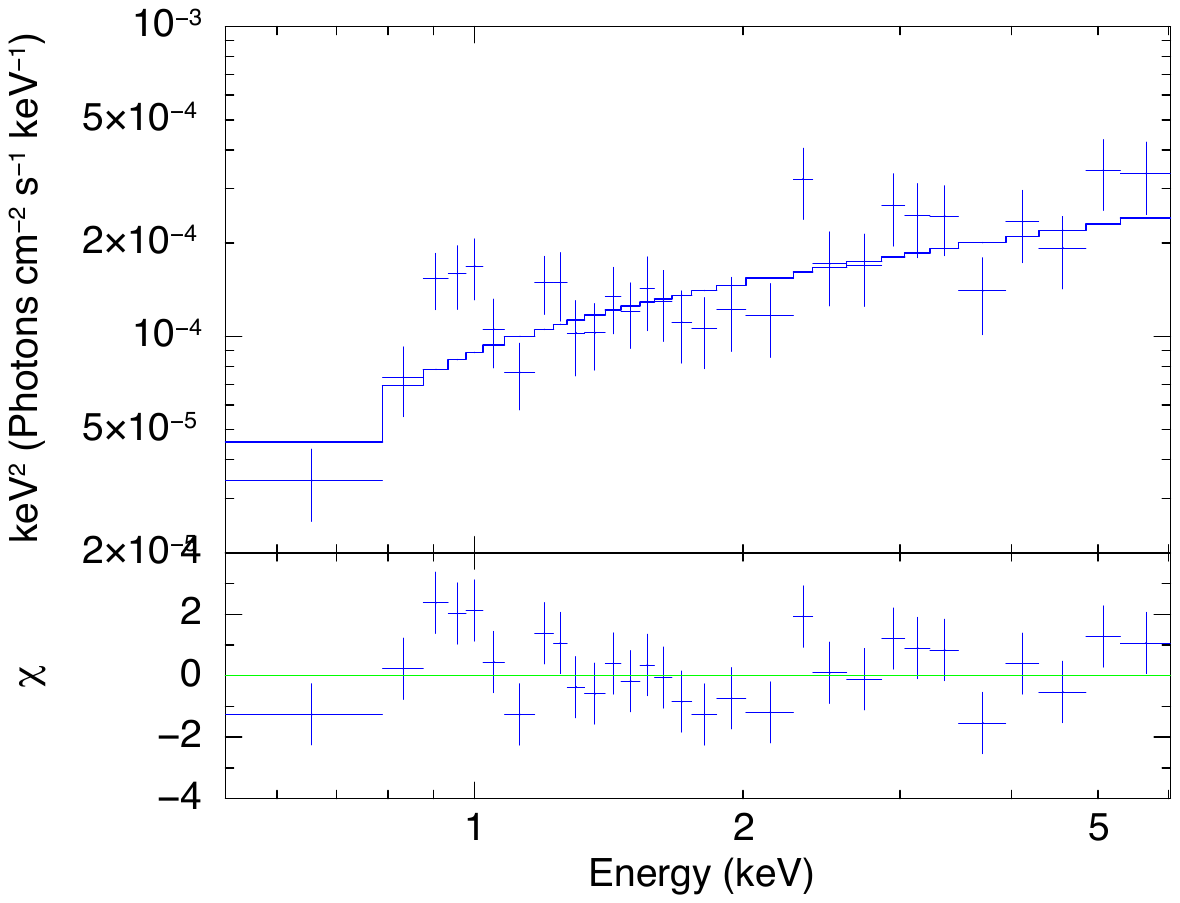} 
\hspace{-2.5cm} 
\vspace{-2cm}
\includegraphics[width=0.45\textwidth, angle=-90]{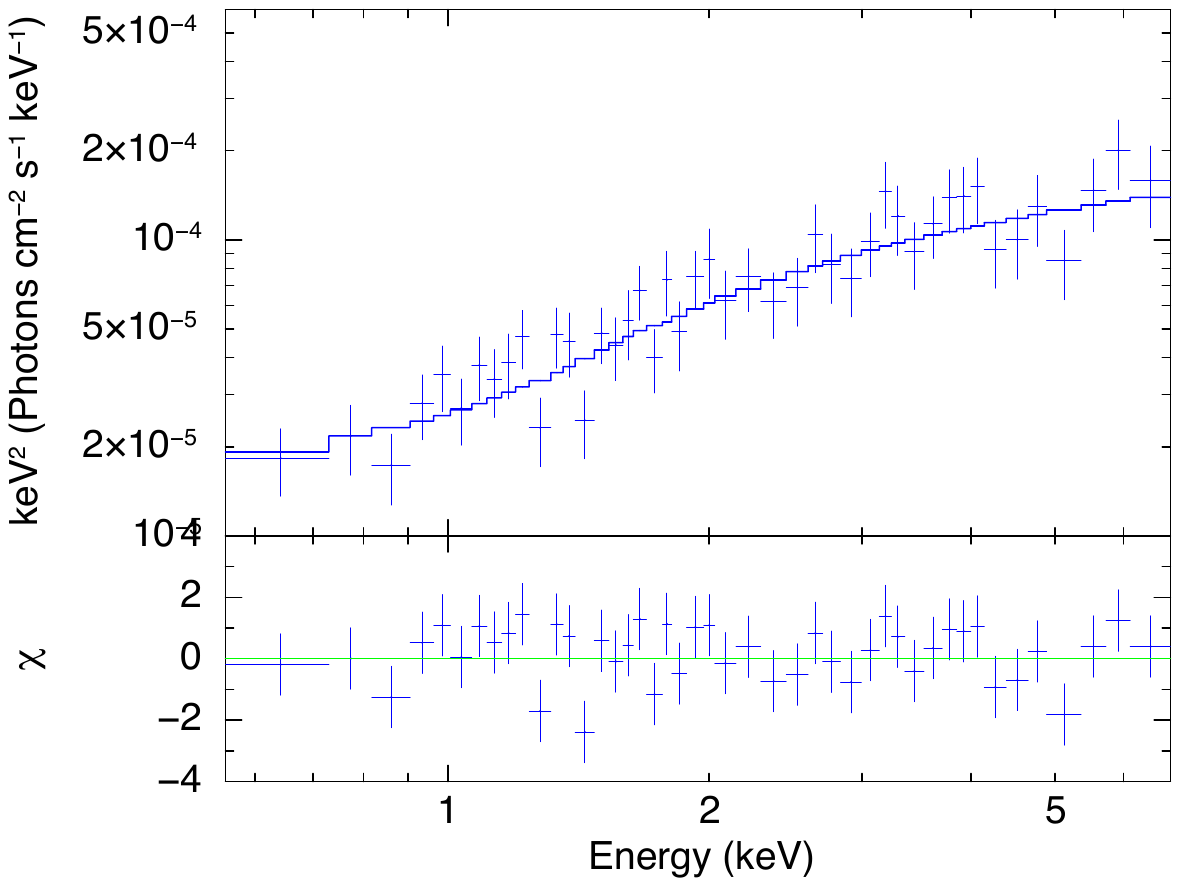}
\vspace{0cm}
\hspace{-1.0cm}
\includegraphics[width=0.45\textwidth, angle=-90]{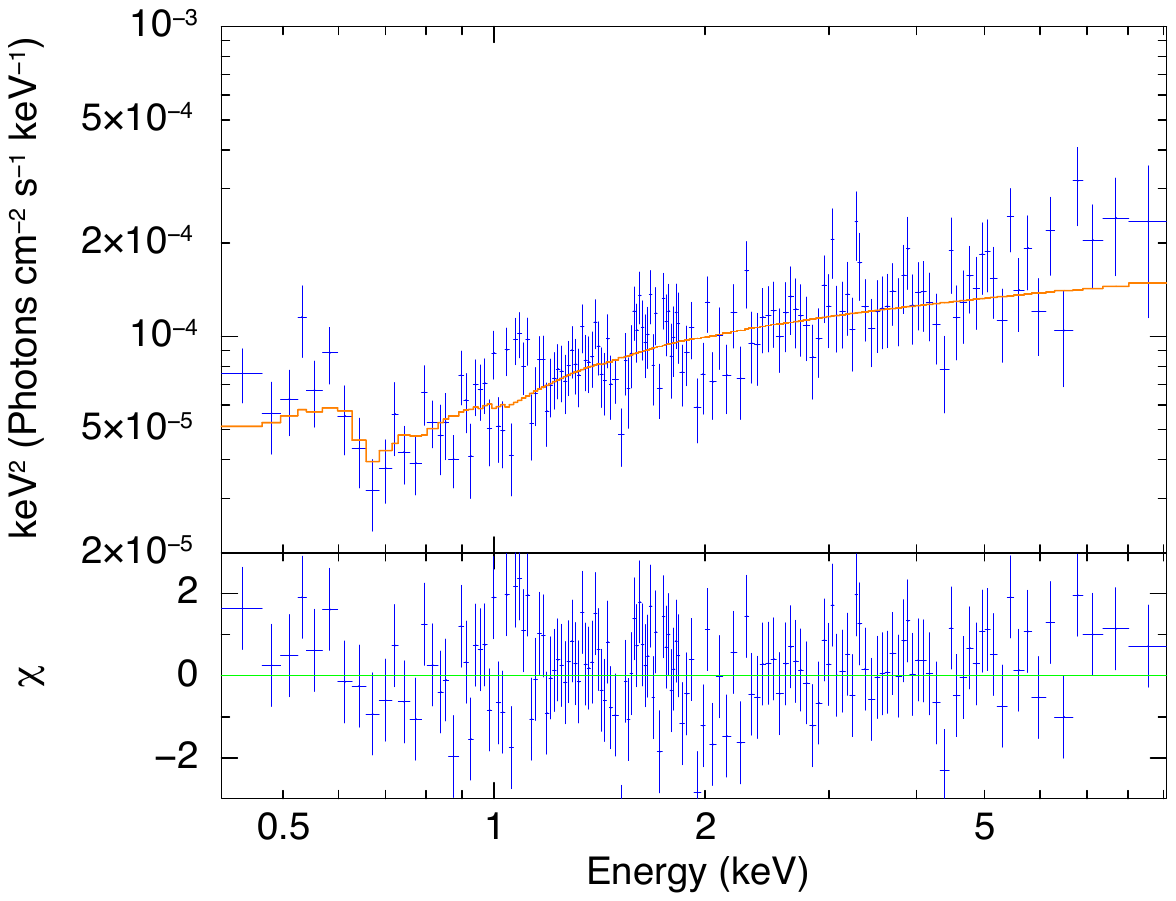} 
\hspace{-2.5cm}
\vspace{-2cm}
\includegraphics[width=0.45\textwidth, angle=-90]{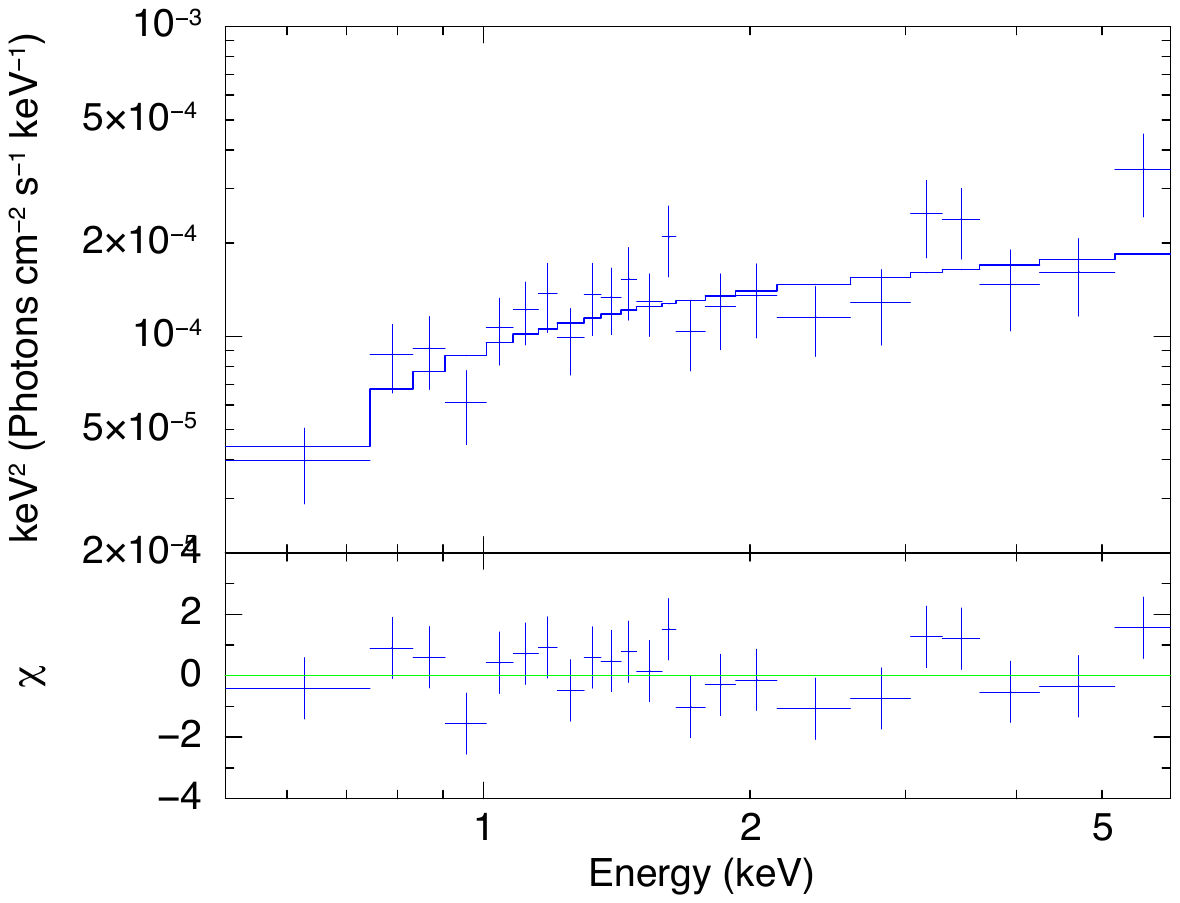} 
\vspace{0cm}
\hspace{-1.0cm}
\includegraphics[width=0.45\textwidth, angle=-90]{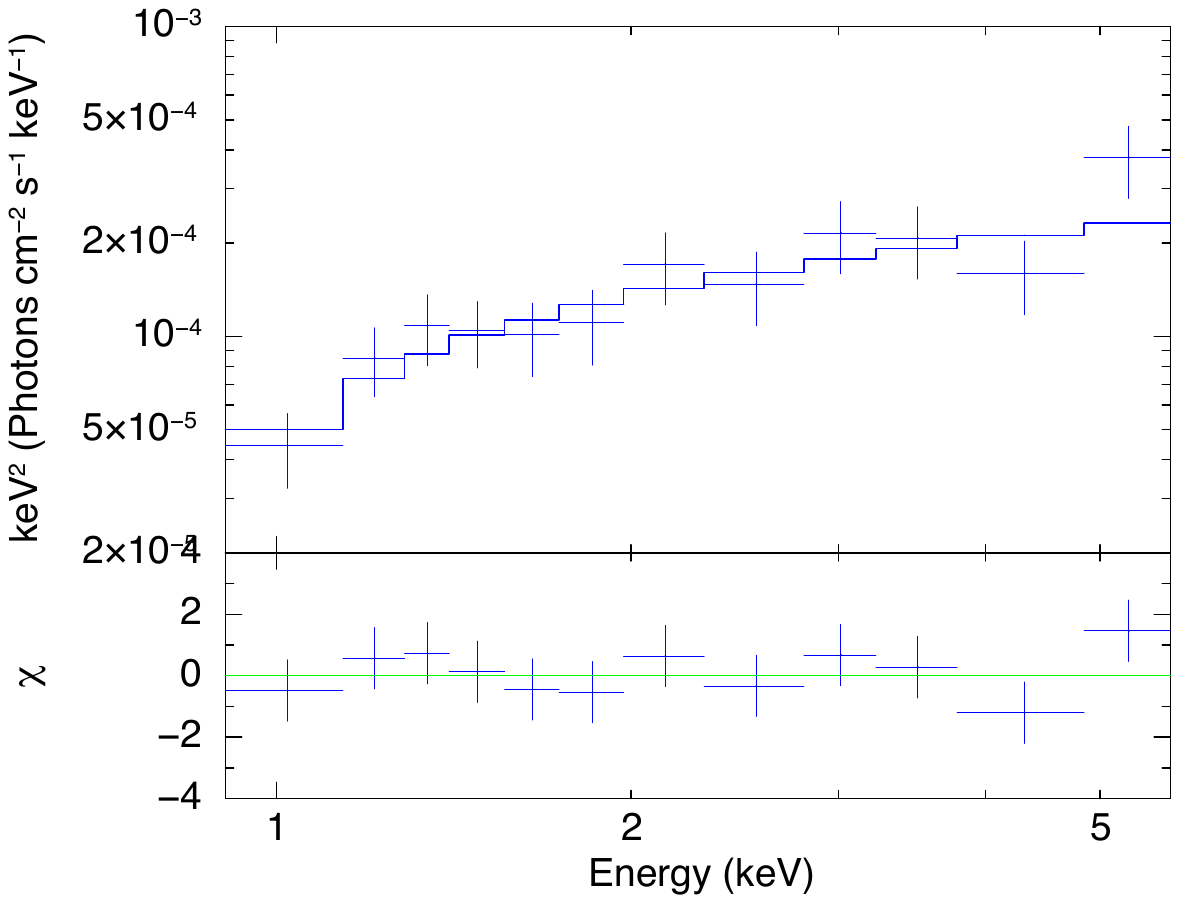}
\vspace{-0.7cm}
\caption{Unfolded spectra for the {\em Chandra} observations 166 (top left), 5002 (top right), 5003 (middle left), 537 (middle right) and 7679 (bottom panel).The spectrum is fitted using the model {\sc Tbabs $\times$ fabsori $\times$ Powerlaw}. }
\label{fig4}
\end{figure*}

Next, we analysed the {\em Chandra} spectra using the same procedure as for {\it XMM-Newton}. All {\em Chandra} spectra are similarly described by the model combination {\sc Tbabs $\times$ fabsori $\times$ (bbody+Powerlaw)}. Given the low data quality of the two observations (Obs\_ID: 537 and 7679), we analysed their spectra jointly in order to improve the signal-to-noise ratio, while displaying the spectra individually. The fitted spectra from the {\em Chandra} observations are shown in Fig.\,\ref{fig4}, and the parameters of the spectroscopic analysis are listed in Table~\ref{tab:chandra_fabsorbi_bb}. We stress that an alternative model for partial covering ionised absorption, such as {\sc zxipcf} would be applicable in the analysis. However, {\sc absori} model was used as it has an index as a parameter that can be tied to the powerlaw index while analysing the X-ray spectrum. Note that {\sc zxipcf} assumes a powerlaw index of 2.2, which is very different from the value we obtained from the actual fit for this source, making them inconsistent. Since {\sc absori} does not come with a partial covering fraction; consequently, we used its variant {\sc fabsori}.

It should also be noted, however, that fitting with the ionised absorber model {\sc fabsori}, either with or without the blackbody component {\sc bbody}, does not yield a substantial difference in terms of the fit statistics. Nevertheless, since the addition of the {\sc bbody} component significantly improves the fit over the absorbed power law alone, we retain both components in order to characterise the soft X-ray band. Just for illustrative purposes, we present the fit results from this model combination in the Appendix (see Tables~\ref{table2} and \ref{table3}).

Apart from the blackbody and the warm absorber components, we investigated whether the power-law model with the inclusion of both Galactic and neutral absorption can improve the fit. Consequently, we use the {\sc zTbabs} model for neutral absorption, so that the complete model becomes {\sc Tbabs $\times$ zTbabs $\times$ Powerlaw}. The inclusion of {\sc zTbabs}, however, does not significantly improve the fit of all spectra, yielding a reduced $\chi^2/{\rm d.o.f.} = 507/294$ (obtained from the {\it XMM-Newton} observation 0502690101). Similar results are obtained for the other observations. In fact, the fit shows strong residuals below 1 keV (as in Figure\,\ref{fig1}), where an absorption-like feature at $\sim0.7$ keV remains. This suggests that neutral absorption alone cannot adequately describe the origin of the soft X-ray band. Furthermore, we examined the fit with the Comptonisation model {\sc compTT} \citep{Titarchuk1994}. In this case, the blackbody component is replaced by the {\sc compTT} model. The fit statistic obtained with this model is not adequately improved, yielding a reduced $\chi^2/{\rm d.o.f.} = 390/292$. We show the fitted spectra from the respective {\it XMM-Newton} observation 0502690101 in Figure\,\ref{bb+comtt+po_fit} (right panel). During the fit, we fix the input seed photon temperature, $kT_{\rm S}$, at 0.01 keV, while leaving the plasma temperature free, which is constrained to be around $<3$ keV. 
It has almost turned out that the phenomenological modelling is adequate to reproduce the observed X-ray spectra.  
An alternate possible scenario, however, is that the contribution to the soft X-ray flux in a fraction could also result from X-ray reprocessing, the so-called X-ray reflection from the disc (or blurred X-ray excess). Consequently, the fit was explored with the reflection model, which did not significantly improve the fit over the absorbed powerlaw model with a warm absorber component. Using the up-to-date relativistic reflection code {\tt relxill} \citep{garcia2013x, garcia2014improved} we obtained a reduced $\chi^2$/d.o.f.=359/292 versus 284/292 using the ionised absorber for {\it XMM-Newton} observation 0502690101. Moreover, the fit could not place a constraint on some of the key parameters, which include the disc inclination $i$, black hole spin $a$, inner emissivity index $q$ reflection fraction $R$, and the iron abundance $A_{\rm fe}$. While this requires a high-quality spectrum, as the model has a larger parameter space, the data used here do not qualify for this. Note further that the phenomenological modelling can describe the spectra well; therefore, more complex reflection modelling was not undertaken in detail, given the data quality. However, we noticed that the inclusion of the warm absorber improves the fit with either of the phenomenological or reflection models, providing a reduced $\chi^2$/d.o.f.=274/284 from the {\tt relxill} fit.

The errors on the parameters are obtained at the 90\% confidence interval by performing an MCMC analysis using the {\sc xspec\_emcee} code\footnote{https://github.com/jeremysanders/xspec\_emcee} developed by Jeremy Sanders. The {\sc emcee} script is a pure Python-based implementation of Goodman \& Weare's Affine Invariant MCMC Ensemble sampler. Here, we used 50 walkers with 10,000 iterations and discarded the first 1,000 as burn-in.

\section{Discussion}
We have produced the first spectroscopic results for the low-redshift quasar 3XMMJ115535.8+232723 using numerous {\it XMM-Newton} and {\it Chandra} observations.

Initial analysis of all datasets has shown that the 0.3--10 keV spectrum imprints a significant absorption feature at $\sim0.7$ keV. This has been accounted for as being due to absorption in the partially ionized matter--the so-called warm absorber, resulting from the ionized {\sc O vii} edge.
The warm absorber column and the spectral index are comparable to those from some of the typical Seyfert galaxies that were found to have warm absorbers. Results obtained for a handful of type 1 Seyfert galaxies are, for instance,  $\Gamma=-1.9$, $N_{\rm H}=5.01\times10^{21}~\rm cm^{-2}$ for NGC 3227 \citep{Komossa1997b}; $\Gamma=-1.9$, $N_{\rm H}=6.3\times10^{21}~\rm cm^{-2}$ for NGC 3786 and $\Gamma=-2.2$, $N_{\rm H}=5.01\times10^{21}~\rm cm^{-2}$ for MGC-6-30-15 \citep{Reynolds1997}.

Two of the {\it XMM-Newton} observations (0551280101/0551280201) taken in 2008 have shown significantly higher ionisation ($\xi  \sim 100/80$) and column density ($N_{\rm H} \sim 12/17 \times 10^{22} $\rm cm$^{-2}$) compared to the other observations, where $\xi \sim 2.64$ and $N_{\rm H} \sim 0.4 \times 10^{22}$\rm cm$^{-2}$. Interestingly, the covering fraction for these two observations is constrained to be $\sim 0.8$. A possible model for these absorptions is that they are produced by clouds close to the central region, which occasionally pass over our line of sight. Since the clouds seem to only partially cover the source, their size must be comparable to the X-ray emitting region. This allows us to make a rough estimate of the distance of the absorber from the central source, $r$. Using the definition for the ionisation parameter \citep{Tarter1969, Krolik1981} as a function of the number density $n_{\rm H}$ and the luminosity $L_{\rm ion}$ of the ionizing source,  
\begin{equation}
    \xi=\frac{L_{\rm ion}}{n_{\rm H}\times r^2} = \frac{L_{\rm ion}}{N_{\rm H}\times r^2} \times \Delta r
    \label{eq1}
\end{equation}
where $\Delta r$ is the size of the cloud in Gravitational radius ($r_{\rm g} = GM/c^2$).
%, which we assume to be about ten times the Gravitational radius ($r_{\rm g} = GM/c^2$). 
In the absence of a direct black hole mass measurement for this quasar, we adopt a fiducial mass of \(M_{\rm BH} \approx 10^8~M_\odot\), which is typical of the quasar population (e.g., 3C\,59; \citealt{Wang2026}; PG\,1115+080; \citealt{Morgan2008}), while an absorber thickness of \(\Delta r \approx 10~R_{\rm g}\) is assumed. We emphasise that both the assumed mass and the assumed absorber thickness are order-of-magnitude estimates

Thus, for a black hole mass of $M = 10^{8}$ solar masses, the distance, $r$, can be estimated to be 
\begin{equation}
    r \sim 1000 \times \Bigl(\frac{\Delta r}{10\,r_{\rm g}}\Bigl)^{1/2}\times \Bigl(\frac{M_{\rm BH}}{{10^8 M_{\odot}}} \Bigl)^{1/2} r_{\rm g}
\end{equation} 
where $r_{\rm g} \sim 1.5 \times 10^{13}$ cms for a $10^8$ solar mass black hole. For the estimates we have used the ionising luminosity in the energy range 0.05--300 keV and the column density listed in Table~\ref{tab:xmm_fabsori_bb}. A similar estimate for the other two XMM-Newton observations with smaller warm absorber column density and ionisation parameter, yields,
\begin{equation}
    r > 50000 \times \Bigl(\frac{\Delta r}{10\,r_{\rm g}}\Bigl)^{1/2}\times \Bigl(\frac{M_{\rm BH}}{{10^8 M_{\odot}}} \Bigl)^{1/2} r_{\rm g}, 
\end{equation}
which is a lower limit since the size of these clouds may be significantly larger than $10 r_{\rm g}$. Thus, a possible scenario is that there are two distinct regions with clouds producing different absorptions. Indeed, AGN sometimes exhibit two phases of the warm absorber, as is evident from the radio-quiet quasar I\,Zw\,1 in which the warm absorber was found to have low- and high-ionisation phases with $\xi \sim 1$ and $ \xi \sim 100$ \citep[see][]{Silva2018}. Consequently, we examined whether the warm absorber in 3XMMJ115535.8+23272 was composed of two ionisation phases in observations 551280101 and 0551280201. 
For these two observations, we included a second {\sc fabsori} component and fixed the covering fraction to $C_f = 1$ and the ionisation parameter to $\xi = 2$, values comparable to those derived from the other two {\it XMM-Newton} observations (0502690101 and 0502690201). This provided a reasonable fit to the spectra, with $\chi^2$/d.o.f. = 44/38 versus $\chi^2$/d.o.f. = 48/39 for a single absorber component in observation 0551280101, and $\chi^2$/d.o.f. = 46/46 versus $\chi^2$/d.o.f. = 51/47 for a single absorber in observation 0551280201 (see Table~\ref{tab:xmm_fabsori_bb}). However, we stress that the interpretation of a two-phase warm absorber remains a plausible scenario, given the fit statistics. 

Using the estimate of the distance to the high density absorbers $r \sim 1000 r_{\rm g}$, one can estimate the typical virial speed of the clouds, $v \sim \sqrt{GM/r} \sim 0.03 c \sim 10^{9} $ cm/s. The cloud crossing time $t_{ct} \sim \Delta r/v$, may then be estimated to be
\begin{equation}
    t_{ct} \sim 2 \times \Bigl(\frac{\Delta r}{10\,r_{\rm g}}\Bigl)^{1/2}\times \Bigl(\frac{M_{\rm BH}}{{10^8 M_{\odot}}} \Bigl)^{1/2} {\rm days}
\end{equation}
This timescale indicates that the variability of the absorber may be on timescales of days. However, this cannot be tested with the available observations used here, which instead allow us to detect variability over the course of observations taken at intervals of $\sim$years.

{\it XMM-Newton} and {\it Chandra} observations of the low red-shift quasar 3XMMJ115535.8+232723 show significant flux and spectral shape variation on time scales of years. The spectra show warm absorber signatures with column densities and ionisation parameters varying by orders of magnitude. The results can be interpreted in a scenario where there are high-density clouds located about $\sim 1000 r_{\rm g}$ from the central source and, much further away, another region of low-density clouds. Future high-resolution spectroscopy of the source will reveal the dynamic nature of these clouds and further constrain the system.

\begin{acknowledgements}
      SH. M. Shehata acknowledges IUCAA for providing comprehensive support for the successful implementation of this research. I also acknowledge the Science, Technology \& Innovation Funding Authority (STDF) and the Institut Français d\'Egypte (IFE) for support through the STDF–IFE postdoc fellowship program (Call 11), in partnership with the Embassy of France in Egypt. SH. M. Shehata also expresses sincere gratitude to the LUX Laboratory, CNRS UMR 8262, Observatoire de Paris, for hosting this research during the postdoctoral fellowship period. SB acknowledges support from China Postdoctoral Science Fund Surface Grant (No. 404985), Shanghai Super Postdoctoral Incentive Plan (No. 2024686), and Ministry of Science and Technology, National Foreign Expert Project (No. 20240238). This research is based on data obtained from the XMM-Newton Science Archive (XSA) and Chandra Data Archive provided by the Chandra X-ray Centre (CXC). Analysis tools and information were collected through the High Energy Astrophysics Science Archive Research Centre Online Service, provided by the NASA/Goddard Space Flight Centre.
\end{acknowledgements}

\bibliographystyle{aa} % style aa.bst
\bibliography{sample701} % your references Yourfile.bib

\appendix
\onecolumn

\section{Spectral parameters from the fits of {\it XMM-Newton} and {\it Chandra} observations using model {\sc Tbabs $\times$ fabsori $\times$ Powerlaw}.}

\begin{table*}[!ht]
\centering
\caption{Results from the spectral fit to the {\em XMM-Newton} dataset using model {\sc Tbabs $\times$ fabsori $\times$ Powerlaw}. Parameters from top to bottom sequentially represent the column density of the warm absorber ($n_{\mathrm{H}}$), ionisation parameter ($\xi_{\mathrm{warm}}$), covering fraction ($C_f$), photon index ($\Gamma$), and  normalisation parameter ($A$)}.
\begin{tabular}{llllllll}
\hline
{\sc Tbabs $\times$ fabsori $\times$ Powerlaw} & \multicolumn{4}{c}{{\em XMM-Newton}} \\
& {0502690101} & {0502690201} & {0551280101} & {0551280201} \\
\hline
$N_{\mathrm{H}} \times 10^{22}$ (cm$^{-2}$) & $0.47^{+0.27}_{-0.05}$ & $0.50^{+0.20}_{-0.06}$ & $12.77^{+5.14}_{-6.30}$ & $14.63^{+4.77}_{-3.67}$ \\\\

$\xi$ (erg cm s$^{-1}$) & $8.33^{+4.51}_{-2.77}$ & $11.46^{+6.02}_{-3.82}$ & $118.88^{+32.35}_{-79.94}$ & $81.68^{+36.28}_{-72.03}$ \\\\

$C_f$ & $\ge0.76$ & $\ge0.82$ & $0.78^{+0.07}_{-0.04}$ & $0.79^{+0.07}_{-0.03}$ \\\\

$\Gamma$ & $2.06^{+0.07}_{-0.05}$ & $1.98^{+0.05}_{-0.05}$ & $1.50^{+0.26}_{-0.15}$ & $1.50^{+0.26}_{-0.153}$ \\\\

$A \times 10^{-4}$ & $2.03^{+0.21}_{-0.26}$ & $1.15^{+0.09}_{-0.15}$ & $0.39^{+1.50}_{-0.15}$  & $0.50^{+0.09}_{-0.23}$\\\\

\hline
$\chi^2/{\mathrm{d.o.f.}}$ & 284/292 & 227/208 & 48/41 & 51/49 \\ 
\hline
\end{tabular}
\label{table2}
\end{table*}
%%%%%%%%%%%%%%%%%%%%%%%%%%%%%%%%%%%%%%%%%%%%%%%%%%%%%%%

\begin{table*}[!ht]
\centering
\caption{Results from the spectral fit to the {\em Chandra} data using model {\sc Tbabs $\times$ fabsori $\times$ Powerlaw}. Parameters from top to bottom sequentially represent the column density of the warm absorber ($n_{\mathrm{H}}$), ionisation parameter ($\xi_{\mathrm{warm}}$), covering fraction ($C_f$), photon index ($\Gamma$), and normalisation parameter ($A$)}.

\begin{tabular}{llllll}
\hline
{\sc Tbabs $\times$ fabsori $\times$ Powerlaw} & \multicolumn{4}{c}{{\em Chandra}}\\
& {1661} & {5002} & {5003} & {537+7696} \\
\hline
$N_{\mathrm{H}} \times 10^{22}$ (cm$^{-2}$) & $0.12^{+0.41}_{-0.09}$ & $2.37^{+3.74}_{-1.28}$ & $0.44^{+0.86}_{-0.23}$ & $0.23^{+5.40}_{-0.01}$ \\\\

$\xi$ (erg cm s$^{-1}$) & $<3.94$ & $0.14^{+83.2}_{-0.25}$ & $16.84^{+25.03}_{-12.22}$ & $3.74^{+48.20}_{-1.68}$ \\\\

$C_f$ & $\ge0.18$ & $\ge0.46$ & $\ge0.50$ & $\ge0.46$ \\\\

$\Gamma$ & $1.57^{+0.25}_{-1.57}$ & $1.60^{+0.46}_{-1.59}$ & $1.79^{+0.13}_{-0.13}$ & $1.62^{+0.75}_{-0.09}$ \\\\

$A \times 10^{-4}$ & $1.15^{+0.25}_{-0.26}$ & $0.93^{+1.40}_{-0.59}$ & $0.96^{+0.23}_{-0.13}$  & $1.08^{+1.83}_{-0.13}$ \\\\

\hline
$\chi^2/{\mathrm{d.o.f.}}$ & 37/25 & 39/40 & 163/143 & $33/29$ \\
\hline
\end{tabular}
\label{table3}
\end{table*}

\end{document}